\documentclass[prb,reprint,superscriptaddress,amsmath,amssymb]{revtex4-2}
\usepackage{bm}
\usepackage{graphicx}
\usepackage{amsmath,amssymb}

\usepackage{float}

\begin{document}

\title{Joule heating effects in high transparency Josephson junctions}

\author{Matti Tomi}
\affiliation{QTF Centre of Excellence, Department of Applied Physics, Aalto University, P.O. Box 15100, FI-00076 Aalto, Finland}
\author{Mikhail R. Samatov}
\affiliation{HSE University, 101000 Moscow, Russia}
\author{Andrey~S.~Vasenko}
\affiliation{HSE University, 101000 Moscow, Russia}
\affiliation{I.E. Tamm Department of Theoretical Physics, P.N. Lebedev Physical Institute, Russian Academy of Sciences, 119991 Moscow, Russia}
\author{Antti Laitinen}
\affiliation{QTF Centre of Excellence, Department of Applied Physics, Aalto University, P.O. Box 15100, FI-00076 Aalto, Finland}
\author{Pertti Hakonen}
\affiliation{QTF Centre of Excellence, Department of Applied Physics, Aalto University, P.O. Box 15100, FI-00076 Aalto, Finland}
\affiliation{Low Temperature Laboratory, Department of Applied Physics, Aalto University, P.O. Box 15100, FI-00076 Aalto, Finland}
\author{Dmitry S. Golubev}
\affiliation{QTF Centre of Excellence, Department of Applied Physics, Aalto University, P.O. Box 15100, FI-00076 Aalto, Finland}

\begin{abstract}
We study, both theoretically and experimentally, the features on the current-voltage characteristic 
of a highly transparent Josephson junction caused by transition of the superconducting leads to the normal state.
These features appear due to the suppression of the Andreev excess current. We show that by tracing the dependence of the voltage, at which the
transition occurs, on the bath temperature and by analyzing the suppression of the excess current by the bias voltage
one can recover the temperature dependence of the heat flow out of the junction.
We verify theory predictions by fabricating two highly transparent 
superconductor-graphene-superconductor (SGS) Josephson junctions with suspended and non-suspended graphene as a non-superconducting
section between Al leads.
Applying the above mentioned technique we show that the cooling power of the suspended junction depends on the bath temperature as
$\propto T_{\rm bath}^{3.1}$ close to the superconducting critical temperature.
\end{abstract}


\maketitle


\section{Introduction}

Highly transparent Josephson junctions possess unique properties
distinguishing them from the usual low-transparency junctions, in which
two superconducting leads are separated by a tunnel barrier. 
For example, transparent junctions have a non-sinusoidal current-phase relation \cite{Haberkorn,Golubov}  
and host Andreev levels, which can be probed by microwave
spectroscopy \cite{Devoret} and can be potentially used for quantum information processing \cite{Shumeiko}.
These junctions also exhibit a characteristic pattern of multiple Andreev reflection (MAR) peaks in the differential conductance
at sub-gap bias voltages $eV<2\Delta$, where $\Delta$ is the superconducting gap \cite{Octavio,Cuevas,Averin,Averin2,Andrei}. 
Two superconducting leads of a transparent Josephson junction
are usually connected by a short section made of a non-superconducting material, which forms good electric contacts to them.
A variety of materials have been used for this purpose:  
carbon nanotubes \cite{Tinkham,CNK,Pillet}, graphene \cite{Calado,Lee}, InAs nanowires \cite{Delsing,Marcus,Devoret}, 
2d \cite{Molenkamp} and 3d \cite{Brinkman,Chalmers} topological insulators, etc.  

Here we consider 
yet another characteristic feature of highly transparent Josephson junctions -- the excess current. 
More specifically, we investigate, both theoretically and experimentally, the suppression of the excess current by Joule heating.
It is known that the I-V curve of a transparent junction at high bias voltage approaches the
asymptotic form $I=V/R + I_{\rm exc}$, where $R$ is the junction resistance in the normal state and
$I_{\rm exc}$ is the excess current. This current is 
proportional to the value of the gap $\Delta$ in the leads and originates from  Andreev reflection \cite{BTK}. 
Joule heating of the superconducting leads suppresses the gap $\Delta$.
At some bias point the temperature of the leads reaches the critical temperature of the superconducting phase transition $T_C$, and
both the gap and the excess current vanish. This bias point is evidenced by a dip in the
differential conductance of the junction. 
Such dips are often observed in Josephson junctions, see e.g. Refs. \cite{Nguyen,Xiong,Andrei}.
They have been systematically investigated by Choi {\it et al} \cite{Choi} in highly 
transparent superconductor - graphene - superconductor (SGS) Josephson junctions on silicon oxide substrate, 
who have made an important observation that the Joule heating power $P_C$, at which
the leads become normal, does not depend on the gate voltage controlling the junction resistance.  
Similar dips in $dI/dV$ are also observed in normal metal - superconductor 
(NS) junctions \cite{Westbrook,Gifford}.
Positions of the dips in the differential conductance provide information 
about temperature dependence of the cooling power, i.e. the heat flow out of the junction, which is important for bolometry applications. 
For example, by tracing dip positions at different bath temperatures $T_{\rm bath}$,
Choi {\it et al} have shown that in an SGS junction with Al/Ti leads on the Si/SiO$_2$ substrate  
$P_C$ approximately scales as $P_C\propto T_C^4-T_{\rm bath}^4$.

In spite of their frequent observation, detailed theoretical description of the dips in $dI/dV$ described above
is still lacking. Historically, several models have been proposed in literature to explain their origin: 
multiple reflections of quasiparticles between the bulk of the superconducting lead and the contact area \cite{Nguyen},
suppression of superconductivity in the leads by high current density \cite{Xiong} and
by magnetic field induced by the current \cite{Westbrook,Gifford}. 
At present, it is understood that Joule heating is the main cause of the above-gap features 
in junctions with the resistance exceeding that of the leads.
The effect of Joule heating on the positions of the dips has been discussed, for example, in Refs. \cite{Westbrook,Choi}.
In a broader context, it has been shown that Joule heating  
suppresses the gap in the leads of a the superconductor - normal metal - superconductor (SNS) junction \cite{Flensberg}, 
induces hysteresis in the current - voltage characteristics of SNS junctions \cite{Pekola} and
of superconducting nanowires \cite{Tinkham2}, etc. 

Here we develop a theoretical model of Joule heating in high transparency Josephson junctions
and derive a simple analytical expression for the differential conductance in the vicinity
of a high bias dip.  We demonstrate that the critical power $P_C$ predominantly
depends on the properties of the leads and the substrate rather than on the junction resistance.
We also test the theory predictions 
experimentally by studying highly transparent SGS Josephson junctions
with aluminum leads. We show, in particular, that for a suspended SGS junction
the cooling power scales as $P\propto T_0^{3.1}-T_{\rm bath}^{3.1}$, where $T_0$ is the temperature
of the superconducting leads at the contacts with graphene.
Graphene is known to be a very promising material for bolometry \cite{Vora_2012,Fong_2012,Yan_2012,McKitterick_2013,Efetov_2018,Han_2013,Cai_2014,Fatimy_2016}.
Recently, bolometers based on SGS junctions similar to ours and having the noise equivalent power at the level $10^{-20} - 10^{-18}$ W/$\sqrt{\rm Hz}$ in
the microwave frequency range have been demonstrated \cite{Lee_2020,Kokkoniemi}. 
We argue that the technique based on the suppression of the excess current by Joule heating may provide
an additional tool for calibration of such bolometers because it allows one 
to obtain the dependence of the cooling power on temperature for a given device in a certain temperature range.

The paper is organized as follows: in Sec. \ref{theory} we present the theoretical model,
in Sec. \ref{experiment} we discuss the experiment and in Sec. \ref{conclusion} we summarize our results.

\section{Model} \label{theory}

In this section we present the theoretical model.
We assume that the bias voltage $V$ applied to the junction is positive and  sufficiently large,  $eV> 2\Delta$.
In this limit the I-V curve of a transparent Josephson junction can be well approximated as
\begin{equation}
I(V) = \frac{V}{R} + \alpha \frac{\Delta_L(T_L) + \Delta_R(T_R)}{2eR}.
\label{I(V)}
\end{equation}
Here the second term in the right hand side is the excess current and
$\Delta_L(T_L)$ and $\Delta_R(T_R)$ are the superconducting gaps 
in the left and the right leads. They depend on the temperatures of the leads at the contacts with the central section, $T_L$ and $T_R$.
For a short Josephson junction with the distance between the leads, $L_{\rm jct}$, much
shorter than the superconducting coherence length $\xi$, i.e. for $L_{\rm jct}\ll \xi$, 
the pre-factor $\alpha$ is expressed as \cite{Zaitsev}
\begin{equation}
\alpha = \frac{\sum_j\frac{\tau_j^2}{1 - \tau_j} \bigg(1 - \frac{\tau_j^2}{2 \sqrt{1 - \tau_j} (2-\tau_j)} 
\ln{\frac{1+ \sqrt{1-\tau_j}}{1 - \sqrt{1 - \tau_j}}}\bigg)}{\sum_j \tau_j}.
\end{equation}
Here $\tau_j$ is the transmission probability of $j$-th conducting channel and the sums run over all channels. 
The parameter $\alpha$ varies between 0 and $8/3$ and approaches its maximum value
in a highly transparent junction with channel transmissions close to 1. In a tunnel junction with very
small transmissions $\tau_j\ll 1$ one finds $\alpha\ll 1$ and the excess current in Eq. (\ref{I(V)}) vanishes.
Two other important cases - 
a normal metal diffusive wire (SNS junction) and a clean short and wide strip of graphene - 
are characterized by Dorokhov's distribution of transmission probabilities \cite{Dorokhov,Titov}, 
which results in $\alpha = {\pi^2}/{4} - 1 \approx 1.4674$.
Finally, in a junction having a long diffusive section with $L_{\rm jct}>\xi$ the pre-factor $\alpha$ decreases with the length as \cite{Kopu} 
\begin{eqnarray}
\alpha = 2.47\,{\xi}/{L_{\rm jct}}.
\label{diff}
\end{eqnarray}

\begin{figure}
\includegraphics[width=\columnwidth]{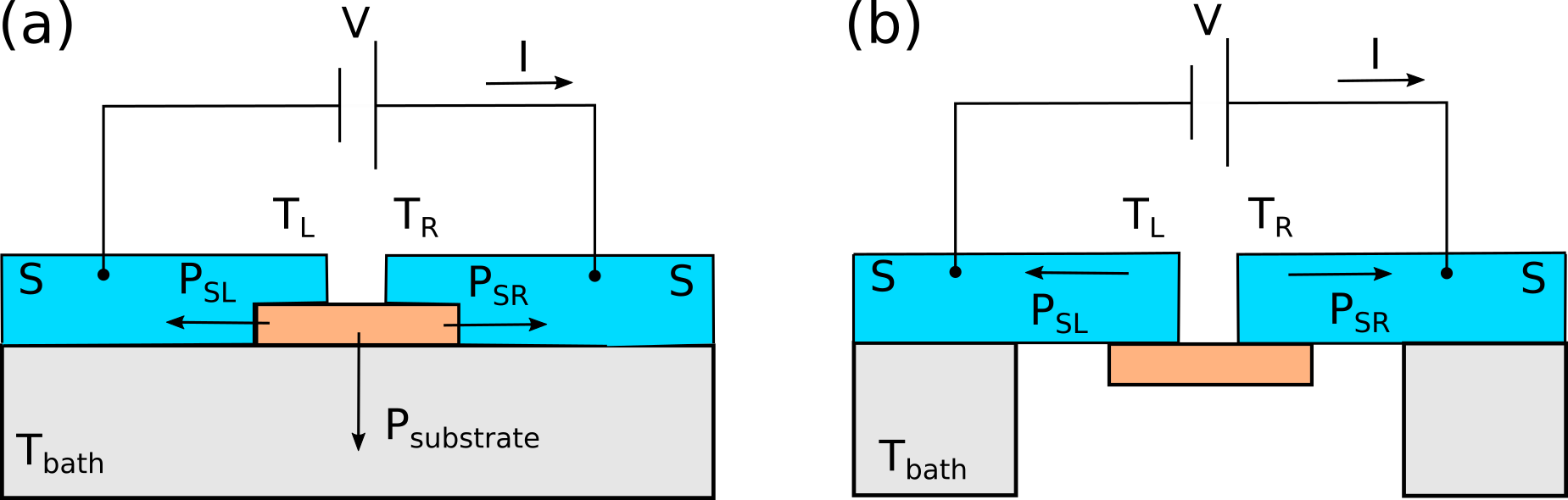}
\caption{(a) High transparency Josephson junction with normal-conductor section connecting the two 
identical superconducting leads, fully residing in contact with the substrate.
Part of the Joule heat goes to the substrate ($P_{\rm substrate}$) and the rest of it -- to the left ($P_{SL}$) and
to the right ($P_{SR}$) superconducting leads.
$T_L$ and $T_R$ are the temperatures of the superconducting leads at the contacts with the central section.
(b) Josephson junction with a suspended central part, in which the Joule heat escapes only through the superconducting leads.}
\label{device}
\end{figure}

We first consider a simple case of  two identical superconducting leads, which are symmetrically coupled to the central normal section. 
Accordingly, we put $T_L=T_R=T_0$ and $\Delta_L(T_L) = \Delta_R(T_R) = \Delta(T_0)$.
In order to find the dependence of the  temperature $T_0$ on the bias voltage $V$ 
we need to solve the heat balance equation
\begin{eqnarray}
P(T_0,T_{\rm bath}) = IV,
\label{heat}
\end{eqnarray}
where we have introduced the bath temperature $T_{\rm bath}$ and the cooling power of the junction $P(T_0,T_{\rm bath})$, which is, in general, unknown.
The Joule heating power is partially carried away to the superconducting leads, while 
the rest of it goes to the substrate phonons and possibly to some
other cooling channels, see Fig. \ref{device}(a). 
Here we assume that the junction resistance is much larger than the resistances of the leads and, hence, the
Josephson critical current of the junction is much lower than the depairing current in the leads. In this case,
at certain bias voltage $V_C$ the contact temperature becomes equal to the critical temperature of the superconductor, $T_0=T_C$.
At this point the differential conductance has a dip. Tracing the position of this dip 
at different bath temperatures, one can restore the cooling power $P(T_C,T_{\rm bath})=I(V_C)V_C$ as a function of $T_{\rm bath}$.

We now find the shape of the dip close to the critical voltage value $V_C$.
In the vicinity of this bias point one can linearize Eq. (\ref{heat}) writing it in the form
\begin{eqnarray}
(T_0-T_C)G_C^{\rm th} = IV-I(V_C)V_C.
\label{linear}
\end{eqnarray}
Here $G_C^{\rm th}=\partial P(T_0,T_{\rm bath})/\partial T_0|_{T_0=T_C}$ is the thermal conductance of
the leads at the critical temperature. 
According to the theory by Bardeen, Cooper and Schrieffer (BCS) \cite{BCS}, 
in the vicinity of the critical temperature the superconducting gap behaves as
\begin{eqnarray}
\Delta(T_0) = k_BT_C\sqrt{\frac{8\pi^2}{7\zeta(3)}\frac{T_C-T_0}{T_C}},
\label{BCS}
\end{eqnarray}
where $\zeta(x)$ denotes Riemann's zeta function.
Combining this expression with the expression for the I-V curve (\ref{I(V)}), we transform Eq. (\ref{linear}) to the form 
\begin{eqnarray}
(T_0-T_C)G_C^{\rm th} = \frac{V^2-V_C^2}{R} + \frac{k_BT_CV}{eR}\sqrt{\frac{8\pi^2\alpha^2}{7\zeta(3)}\frac{T_C-T_0}{T_C}}.
\nonumber
\end{eqnarray} 
Solving this equation for $T_0$, substituting the result in the expression for the current (\ref{I(V)}) and taking
the derivative of it over $V$, we obtain a simple analytical expression for the differential conductance
\begin{eqnarray}
\frac{dI}{dV} = \frac{1}{R} -  \frac{4\theta(V_C-V)}{b R}
\left(\frac{(b-4)V}{\sqrt{b V_C^2 - (b-4)V^2}} + 2\right).
\label{dIdV}
\end{eqnarray}
Here $\theta(x)$ is the Heaviside step function and we have  introduced the dimensionless parameter
\begin{eqnarray}
b = \frac{14\zeta(3)}{\pi^2\alpha^2}\frac{e^2R G_C^{\rm th}}{k_B^2T_C}.
\label{alpha}
\end{eqnarray}
The approximation (\ref{dIdV}) is formally valid if $b\gg 1$ and $ |V_C-V|\lesssim \alpha k_BT_C/e$.
However, Eq. (\ref{dIdV}) can also be used outside this voltage interval because the non-linear correction
to $dI/dV$ almost vanishes there.  
Eq. (\ref{dIdV}) predicts strong dip in the 
differential conductance with the negative minimum value, $-1/R$, which is achieved at $V=V_C$.
In the experiment, the dip is smeared by the inhomogeneity of the hot spot in the contact area between the leads and the
central section, by the inverse proximity effect suppressing the gap close to the contact, etc.

\begin{figure*}[!ht]
\includegraphics[width=1.7\columnwidth]{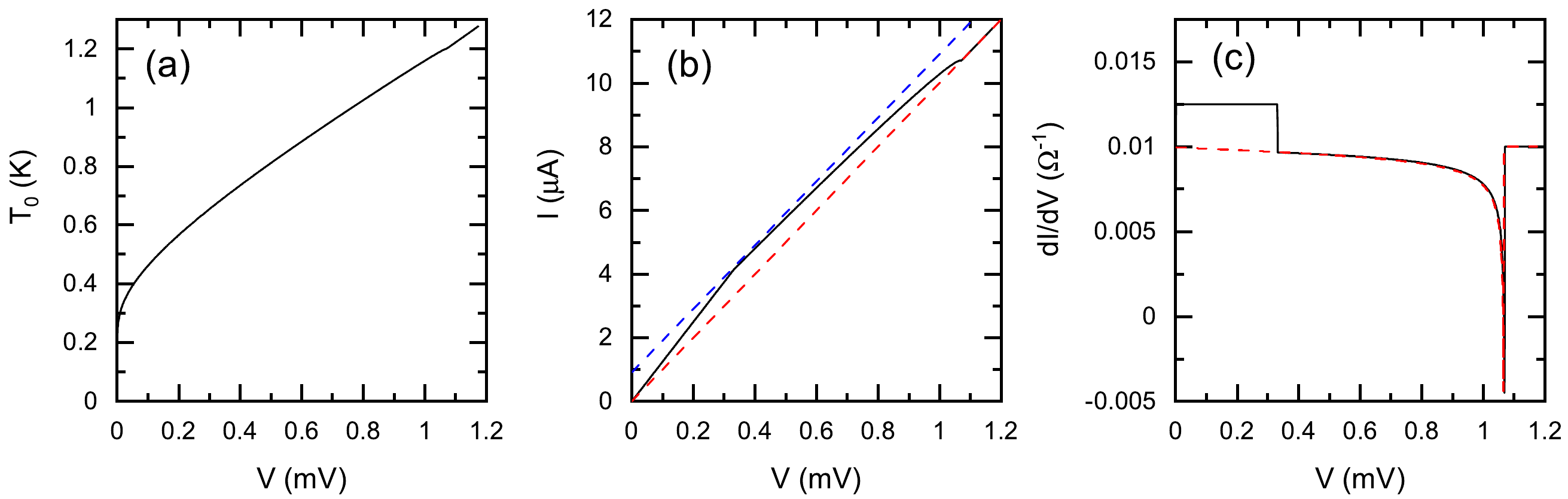}
\caption{(a) Dependence of the contact temperature $T_0$ on the bias voltage.
(b) Solid line is the I-V curve of the junction, 
red dashed line shows Ohm's law $I=V/R$, and blue line -- Ohm's law with added excess current, $I=V/R+\alpha\Delta(0)/eR$. 
The superconducting branch at $V=0$ is not shown. 
(c) Solid line shows the differential conductance  obtianed by solving Eq. (\ref{T(V)_equation}) numerically,
and red dashed line represents the approximate expression (\ref{dIdV}).
The following parameters have been chosen for this simulation: $R=100$ $\Omega$, $R_{L_S}=2$ $\Omega$, $T_C=1.2$ K, 
$\Delta(0)=182$ $\mu$eV, $\alpha=0.5$, $T_{\rm bath}=10$ mK. }
\label{plots}
\end{figure*}

In reality the two superconducting leads of the junction are never fully identical.
They may have different critical temperatures, different geometry,
and different couplings to the central section.
As a result, a single dip in $dI/dV$ splits into two dips occurring at different bias voltages $V_{CL}$ and $V_{CR}$, 
at which the left and the right leads switch to the normal state.
In this case Eq. (\ref{heat}) cannot be solved analytically. However, in the limit $b\gg 1$ one can rather accurately approximate
the solution by the sum of two independent dips,
\begin{eqnarray}
\frac{dI}{dV} &=& \frac{1}{R} -  \frac{2\theta(V_{CL}-|V|)}{b R}\left(\frac{(b-1)|V|}{\sqrt{b V_{CL}^2 - (b-1)V^2}} + 1\right)
\nonumber\\ &&
-\,  \frac{2\theta(V_{CR}-|V|)}{b R}\left(\frac{(b-1)|V|}{\sqrt{b V_{CR}^2 - (b-1)V^2}} + 1\right).
\label{asym}
\end{eqnarray}
For a symmetric case $V_{CL}=V_{CR}$ and for $b\gg 1$ Eq. (\ref{asym}) coincides with the exact Eq. (\ref{dIdV})
everywhere except a narrow interval close to the minimum of the dip.
Eq. (\ref{asym}) is valid for both positive and negative bias and
it assumes that the two dips are sufficiently close so that one can use the same parameter $b$ (\ref{alpha}) for both of them.

\subsection{Suspended symmetric junction}
\label{suspended}

In this subsection we consider a suspended junction in a symmetric configuration, in which the only cooling mechanism is
the removal of heat through the superconducting leads, see Fig. \ref{device}b. This model can be easily solved
numerically, and it allows us to test the analytical approximation (\ref{dIdV}). 
By symmetry, the power $IV/2$ is dissipated in each lead. We assume that the right lead
has the shape of a one-dimensional wire with the cross-sectional area $S$, although the final 
expression for the cooling power of the junction (\ref{PSfull}) given below, 
does not depend on geometry of the lead thanks to the Wiedemann–Franz law relating heat and electric currents. 
Then, the temperature profile in the right lead is determined by the heat diffusion equation
\begin{equation}
    S\kappa_S(T)\frac{dT}{dx} = -\frac{IV}{2},
    \label{balance_equation}
\end{equation}
where  the themal conductivity of the lead is \cite{Bardeen},
\begin{equation}
    \kappa_S(T) = \frac{4k_B^2\sigma}{e^2} T \int^\infty_{\frac{\Delta(T)}{2k_BT}} \frac{x^2}{\cosh^2 x} dx.
    \label{ks(T)}
\end{equation}
Here we have introduced the conductivity of the superconducting material in the normal state $\sigma$. 
We assume that the contact with the non-superconducting central part is located at $x=0$ so that $T(0)=T_0$, 
and at $x=L$ the superconducting wire contacts the bulk lead kept at the bath temperature, $T(L)=T_{\rm bath}$.
Integrating Eq. (\ref{balance_equation}) from $0$ to $L$, we obtain the equation for the contact temperature $T_0$,
\begin{eqnarray}
P_{S}(T_0,T_{\rm bath})  =  I(V)V,
\label{T(V)_equation}
\end{eqnarray}
where 
\begin{eqnarray}
P_{S} = \frac{8k_B^2}{ e^2R_{\rm leads} }\int_{T_{\rm bath}}^{T_0} dT\,T \int_{\frac{\Delta(T)}{2k_B T}}^\infty dx \frac{x^2}{\cosh^2 x}
\label{PSfull}
\end{eqnarray}
is the power carried away by quasiparticles through both superconducting leads.
In Eq. (\ref{T(V)_equation}) we introduced the total resistance of the two leads in the normal state $R_{\rm leads} = 2L/\sigma S$. 
In this model, the heat conductance at critical temperature is determined by Wiedemann–Franz law,
$G_C^{\rm th}=2\pi^2 k_B^2T_C/3e^2 R_{\rm leads}$, and the dimensionless parameter (\ref{alpha}) acquires the form
\begin{eqnarray}
b = \frac{28\zeta(3)}{3\alpha^2}\frac{R}{R_{\rm leads}}.
\end{eqnarray}
Next, from Eq. (\ref{T(V)_equation}) we find the value of the critical voltage $V_C$ at low bath temperatures $T_{\rm bath}\ll T_C$, 
\begin{eqnarray}
V_C\approx \sqrt{ RP_{S}(T_C,0)} = 0.829 \sqrt{\frac{R}{R_{\rm leads}}} \frac{\Delta(0)}{e},
\label{Vc}
\end{eqnarray}
where we have numerically evaluated the cooling power $P_{S}(T_C,0)$  assuming the BCS dependence $\Delta(T)$. 
Assuming that the excess current is fully suppressed for $V>V_C$, 
one finds the critical power as $P_C=V_C^2/R$, which gives
\begin{eqnarray}
P_C= P_S(T_C,0)=0.687\left({\Delta^2(0)}/{e^2 R_{\rm leads}}\right).
\label{Pc}
\end{eqnarray}  

Interestingly, Eqs. (\ref{Vc},\ref{Pc}) agree with many experimetal observations reported in the literature. Indeed, 
experiments in finite magnetic field \cite{Xiong} have shown that the critical voltage
is proportional to the superconducting gap, $V_C\propto \Delta(0)$,  
and the experiments with tip-shaped superconducting electrodes \cite{Westbrook,Gifford} demonstrated  that it is proportional to
the square root of the interfacial resistance $R_{IF}$ ($V_C\propto\sqrt{R_{IF}}$). 
Eq. (\ref{Pc}) predicts that the critical
power does not depend on the junction resistance in agreement with the experiment \cite{Choi}.
We also note that at sufficiently low temperatures Eq. (\ref{Vc}) may also be applicable to junctions lying on the substrate if
the overheated leads have much higher heat conductance than the substrate.

In Fig. \ref{plots} we present numerically exact results obtained from Eq. (\ref{T(V)_equation}).
We have chosen $\alpha=0.5$ and approximated the I-V curve as follows: $I(V) = (1+\alpha/2)V/R$ for $eV<2\Delta$, and
$I(V)=(V + \alpha\Delta/e)/R$ for $eV>2\Delta$. We have ignored MAR features at sub-gap voltages because
we focus at voltages above the double gap,  $eV>2\Delta$.
Voltage dependence of the contact temperature $T_0(V)$
and of the current $I(V)$ are shown in Figs. \ref{plots}(a) and \ref{plots}(b) respectively. 
In Fig. \ref{plots}(c) we compare the exact differential conductance with the approximate expression (\ref{dIdV}) and
observe perfect agreement between them for all voltages $eV>2\Delta$.

Finally, we provide approximations for the quasiparticle cooling power (\ref{PSfull}) at low and at high temperatures.
Eq. (\ref{PSfull}) can be re-written in the form
\begin{eqnarray}
P_S = F(T_0)-F(T_{\rm bath}),
\end{eqnarray}
where at low temperatures $k_BT\ll \Delta(0)$ the function $F(T)$ is exponentially small,
\begin{eqnarray}
F(T) = \frac{4\Delta(0) k_BT}{e^2 R_{\rm leads}}  e^{-{\Delta(0)}/{k_BT}},
\label{PS_low}
\end{eqnarray}
while at temperatures slightly below $T_C$ and for $T>T_C$ it takes the form
\begin{eqnarray}
&& F(T) = \frac{\pi^2 k_B^2T^2}{3e^2 R_{\rm leads}}
\nonumber\\&&
-\, \frac{2}{15}\left(\frac{8\pi^2}{7\zeta(3)}\right)^{3/2}\frac{k_B^2T_C^2}{e^2R_{\rm leads}}
{\rm Re}\,\left[\left(1-\frac{T}{T_C}\right)^{{5}/{2}} \right].
\label{PS_high}
\end{eqnarray} 
The first term in the right hand side of Eq. (\ref{PS_high}) corresponds to the Wiedemann–Franz heat current through the normal leads, and the second term
describes a small correction to it caused by superconductivity onset at temperatures below $T_C$.


\section{Experiment}
\label{experiment}

Our back-gated devices were fabricated using exfoliated graphene on a SiO$_2$/Si wafer. The substrate was highly boron-doped Si (p$^{++}$/B, $\rho < 10$ m$\Omega$cm) which remains conducting down to mK temperatures. A layer of dry oxide with a thickness of $250 - 270$\,nm was grown thermally on the substrate  at a temperature of 1000$^{\circ}$C. Before depositing graphene, alignment markers were patterned on the chips using optical lithography and evaporation of Ti and Au. 

For suspended samples, we made graphene exfoliation on LOR resist \cite{Tombros2011}. After making the electrical contacts using electron beam lithography, the area to be suspended was irradiated using a high dose of electrons, developed in ethyl lactate, and washed in hexane. The electrical leads were made of aluminum/titanium sandwich structures. Titanium was evaporated at ultra high vacuum conditions which resulted in high transparency contacts to graphene. The presence of Ti lowered the superconducting transition temperature of the leads: in our second sample structure with a trilayer sandwich structure 10/50/10\,nm of Ti/Al/Ti, the critical temperature was suppressed to half of  the bulk value. The normal state resistance of the Al leads is $< 1$\,$\Omega/\mu$m, which means that the Joule heating by the measurement current will mostly take place in the SGS junction. For details of the structure and dimensions of the two samples see Table I.

\begin{figure}[b]
\includegraphics[width=0.9\columnwidth]{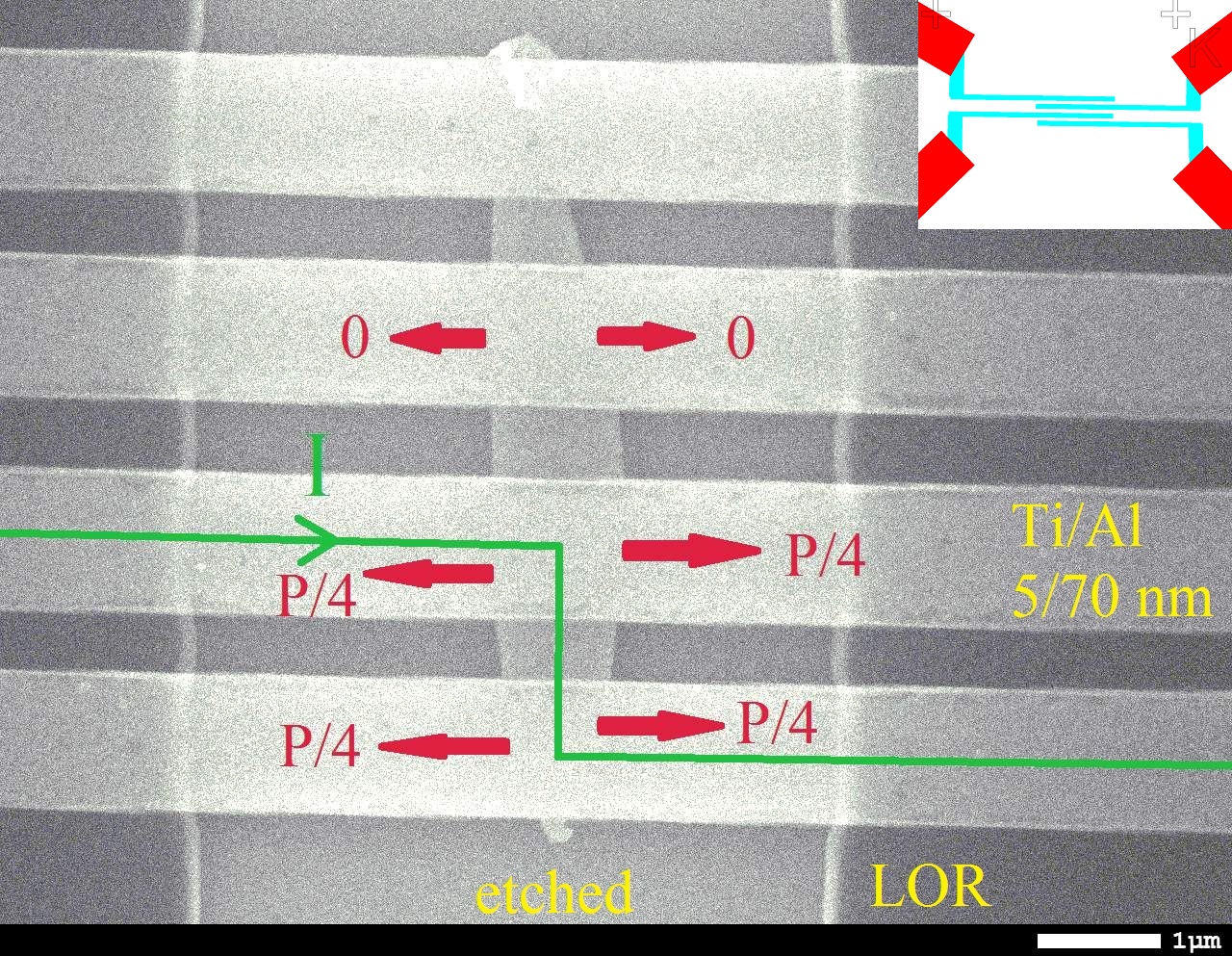}
\caption{Samples structure with the suspended (etched) part visible in the center between the vertical borders of the LOR resist seen as the darker background color. The Al/Ti measurement leads are seen as horizontal lighter grey stripes, the connection of which to the measurement system is displayed in the upper right corner. The green arrow depicts the electrical current flow in our experiments, while the red arrows illustrate the heat flows to the metallic leads as assumed in our analysis.} \label{fig:sample}
\end{figure}

\begin{table}[t]
    \begin{ruledtabular}
    \begin{tabular}{ccccccccc}
         &type& $L$ &$W$&leads &$d$ (nm)&$W_{\mathrm{lead}}$&$N$&$T_C$ (K)\\
         \hline
      NS   & ML & 0.30 & 6.0 & Ti/Al/Ti& 10/50/10&0.6&2&0.58\\
      S   & BL & 0.50 & 1.0 & Ti/Al& 10/70&1.2&4&0.77\\
    \end{tabular}
    \caption{Parameters of the two studied samples: NS - nonsuspended, S - suspended; the types ML and BL refer to monolayer and bilayer graphene, respectively. $L$ and $W$ denote the sample length and width in micrometers, respectively. Column $d$ indicates the thickness of the metal layers evaporated for the contacts, while $W_{\mathrm{lead}}$ denotes the lead width in $\mu$m and $N$ is the effective number of  leads for hat transport. Critical temperature $T_C$ represents an average value of the superconducting transition temperatures observed on different leads.}
    \label{tab:parameters}
    \end{ruledtabular}
\end{table}

The suspended sample (S in Table I) with four measurement leads is illustrated in Fig. \ref{fig:sample}. The width of measurement leads is 1.2\,$\mu$m, i.e. the contact areas are more than two times larger than the size of the actual suspended part of the graphene. The bright looking sections of the horizontal leads in Fig. \ref{fig:sample} are also suspended. Thus, the Joule heating generated in graphene has to propagate along suspended metallic leads over a distance of at least two microns  before any cooling by the substrate through the LOR resist may take place. The arrows in the figure illustrate the flow of electrical current as well as the main directions of Joule heat escape from the sample. The critical temperature was slightly different for the four leads and the obtained $T_C$ values varied between $T_C=760-780$\,mK (expected BCS gap $\Delta=115 \dots 118$\,$\mu$eV).

\begin{figure}[b]
\includegraphics[width=0.9\columnwidth]{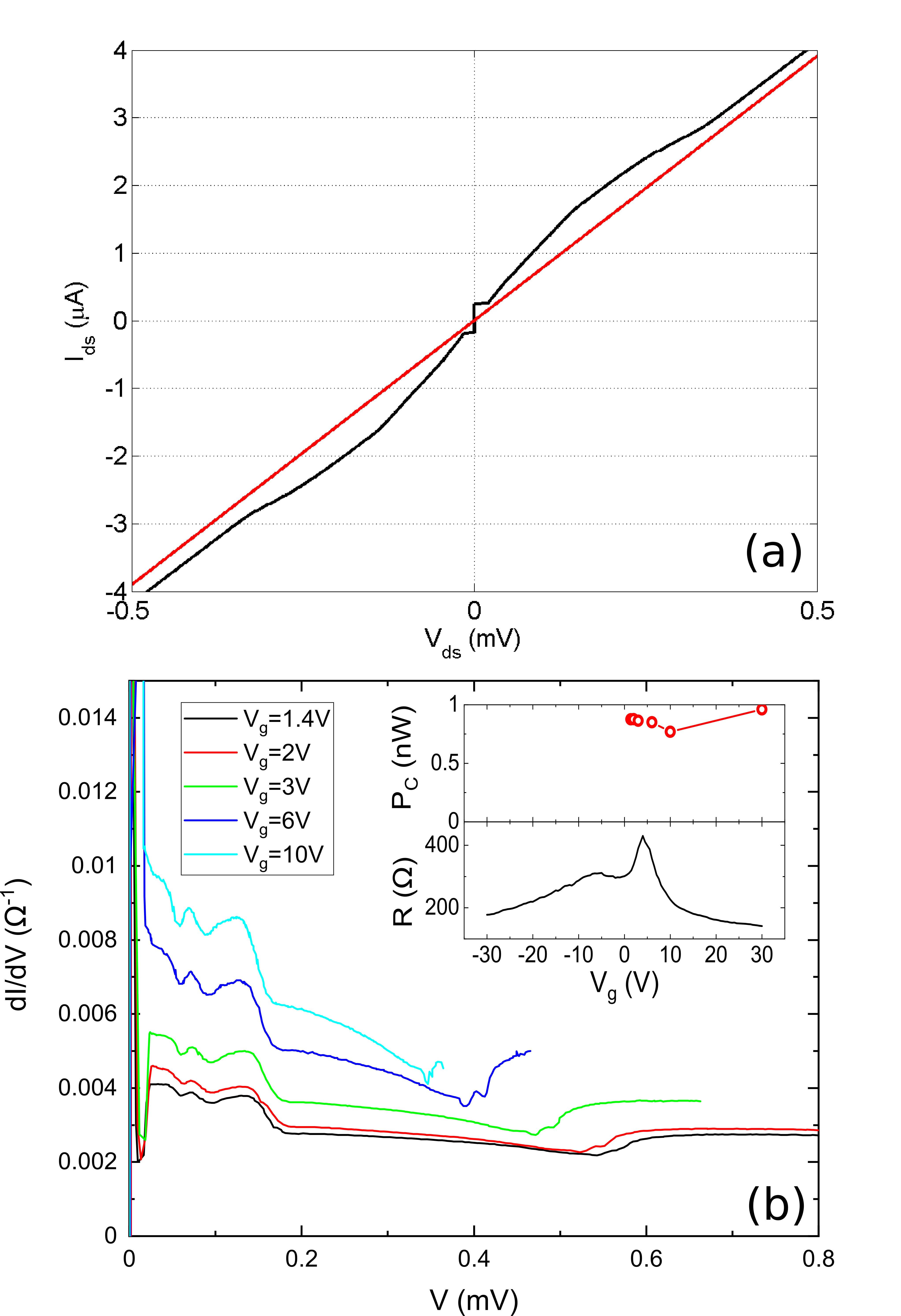}
\caption{(a) Measured $IV$ curve of the sample NS at gate voltage $V_g=30$\,V 
(electron concentration $n \simeq 2 \times 10^{12}$\,cm$^{-2}$) with the normal state resistance $R=128$\,$\Omega$ (indicated by the red line). Parameters of the sample: 
excess current above the double gap voltage $I_{\rm exc}^{(1)}=455$ nA, excess current at high bias $I_{\rm exc}^{(2)}=216$ nA,
$\Delta=87.5$\,$\mu$eV, the parameter $\alpha=1.5$,
critical current $I_C=260$\,nA, re-trapping current $I_r=175$\,nA.
(b) $dI/dV$ at several values of $V_g$. The inset shows the critical power $P_C$ (red circles) and the normal state resistance $R$ (black line)
as functions of $V_g$.}
\label{NS_IV}
\end{figure}

\begin{figure*}[t]
\includegraphics[width=1.7\columnwidth]{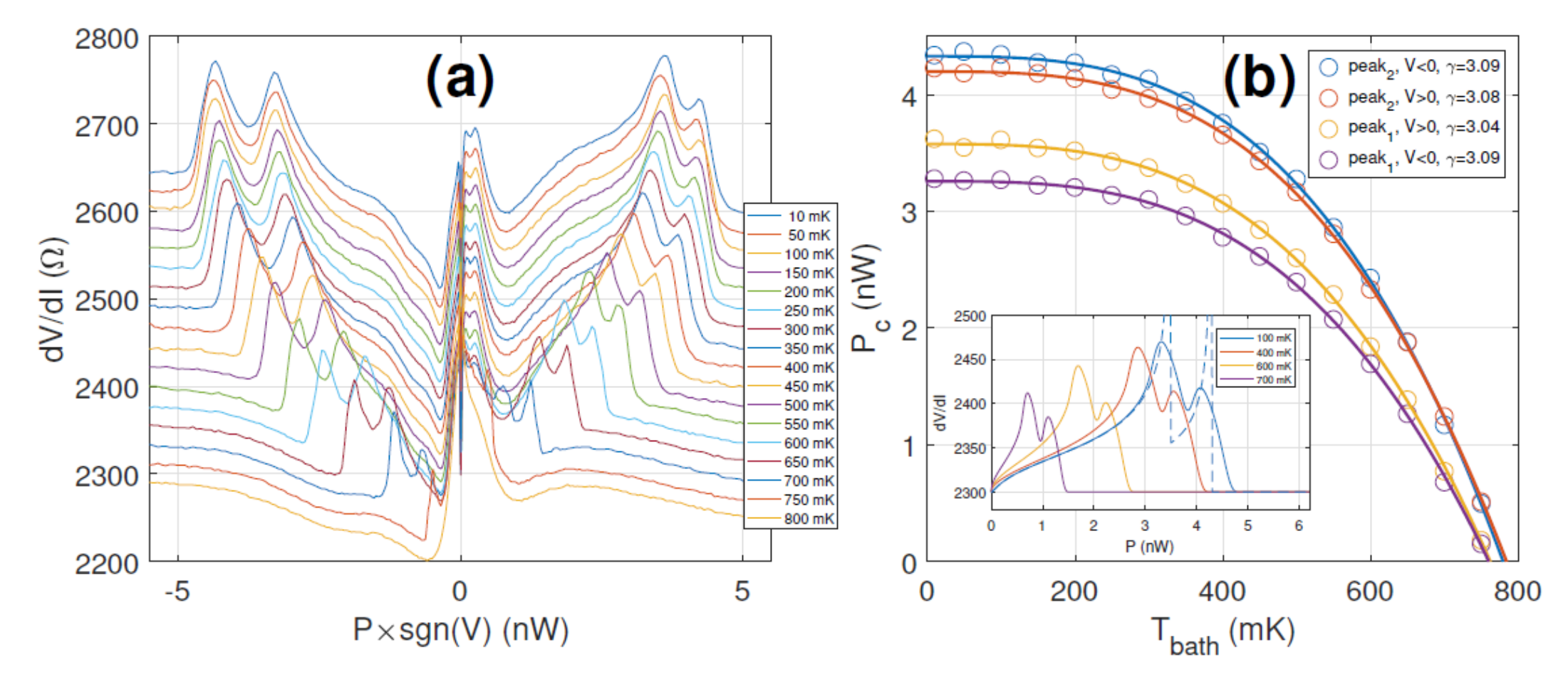}
\caption{Heating effects in a suspended bilayer SGS sample with the following parameters: 
normal state resistance $R=2.2$\,k$\Omega$, maximum excess current $I_{\rm exc}=26.5$\,nA, superconducting gap
$\Delta\approx 90$\,$\mu$eV, $\alpha=0.646$.
(a) Differential resistance as a function of power  at different bath temperatures 
(successive curves have been shifted vertically by 20\,$\Omega$.
For clarity, we define the power as $P=IV\times{\rm sign}V$, so that it has different signs for positive and negative bias voltages. 
(b) Critical powers $P_C$, i.e. positions of the resistance peaks in frame \ref{heating}(a), as a function of bath temperature. Continuous lines are fits using the heat flow power law in Eq. (\ref{PC}) with the exponent $\gamma$
 as a fitting parameter. All these data were measured at $V_g = -20$\,V which corresponds to a hole carrier density of $n=2 \times 10^{11}$\,cm$^{-2}$. Inset: Simulated results for the superconducting-to-normal transition due to Joule heating, where the resistance peaks arise from a reduction of excess current. Solid lines have been smoothed by a moving average, whereas the
dashed blue line displays the unsmoothed result for $T_{\mathrm{bath}} = 100$\,mK.}
\label{heating}
\end{figure*}

Measurements were carried out using two dilution refrigerators: a BlueFors prototype cryostat BF-SD125 and a commercial BlueFors BF-LD250. Most of the measurements were performed on the BF-LD250 cryostat with the lowest achievable base temperature of approx. 10\,mK. Temperatures were measured with a RuO$_2$ resistor that had been calibrated against a Coulomb blockade thermometer \cite{Pekola1994}. An electrically shielded room with filtered power lines was employed to prevent electromagnetic interference and low-frequency electrical noise from aﬀecting the weak signals before ampliﬁcation. To limit the amount of thermal noise entering from room temperature to the sample, all dc lines were equipped with three-stage $RC$ low-pass filters at the base temperature of the cryostat, with a cut-off frequency of $f_{co} \simeq 1$\,kHz.

The electrical characteristics were obtained using regular low-noise measurement methods and lock-in techniques. We employed current bias through $1-100$\,M$\Omega$ resistor, selected on the basis of the required current range (dependent on gate voltage $V_g$). The voltage across the sample was measured using an LI-75 low-noise preamplifier powered from lead batteries. Differential resistance at small audio frequency (around 70\,Hz) was measured simultaneously with the DC characteristics using a Stanford SR830 lock-in amplifier. 

In Fig. \ref{NS_IV}(a) we show the I-V curve of the non-suspended SGS junction taken at low bath temperature $T_{\rm bath}=90$ mK.
This junction has rather large excess current exceeding
the Josephson critical current. For this reason, its suppression  at high bias voltage is clearly visible.
The experimental I-V curve resembles the theoretical one plotted in Fig. \ref{plots}(b), although in the experiment
only one of the leads completely switches to the normal state in presented range of bias voltages. This junction exhibits large value
of the pre-factor in the excess current, $\alpha=1.5$, which points to its high transparency.
Assuming that the junction is long and diffusive, we can use Eq. (\ref{diff}) and estimate the ratio
$L_{\rm jct}/\xi \approx 1.65$, which gives the coherence length $\xi\approx 180$ nm. 
In Fig. \ref{NS_IV}(b) we plot the differential resistance of the same sample at several
values of the gate voltage. It is evident that both the critical voltage $V_C$ and the junction resistance $R$ (see the inset) 
significantly vary with $V_g$.
However, as the inset shows, the critical power $P_C$ only weakly depends on $V_g$ and varies from 0.77 nW to 0.96 nW
for the presented five values of the gate voltage. 
This obervation agrees with that of Ref. \cite{Choi},
and it is also consistent with Eqs. (\ref{Vc},\ref{Pc}). 
We believe that the remaining weak dependence of $P_C$ on $V_g$ results from the incomplete suppression of
the $V_g$-dependent excess current above $V_C$ in this sample.

\begin{figure*}
\includegraphics[width=2\columnwidth]{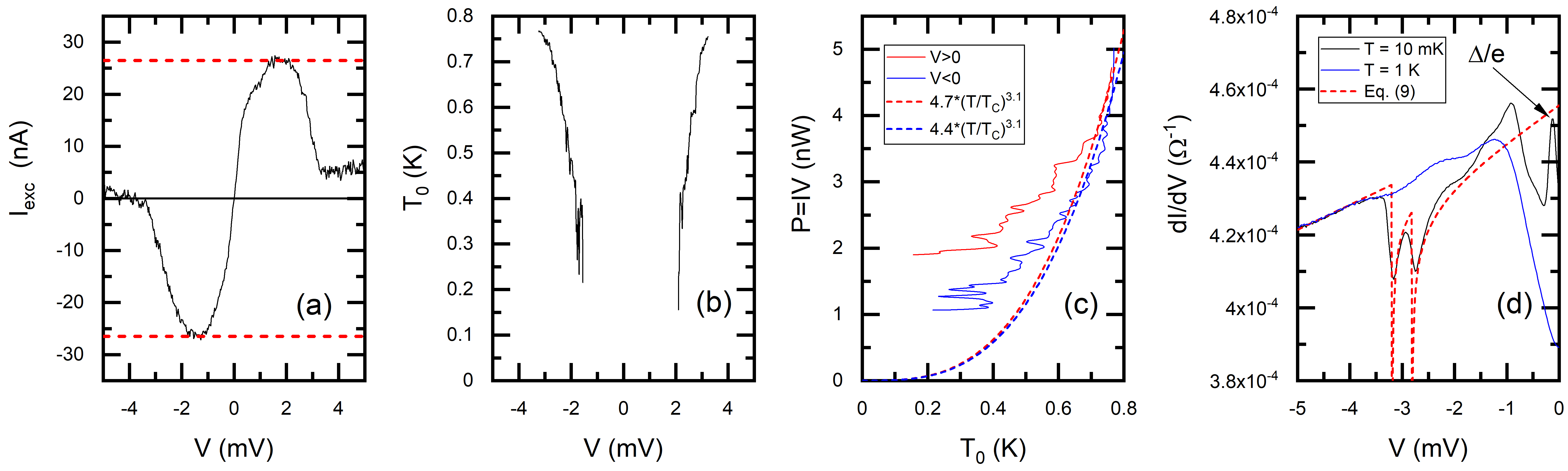}
\caption{(a) Excess current versus bias voltage in a suspended SGS junction measured at the base temperature ($T=10$ mK).
(b) Temperature of the leads in the vicinity of the suspended SGS junction, $T_0(V)$.
(c) Cooling power of the leads versus the temperature at the graphene/Al contacts (given in frame (b)) for the positive (red line) and for the negative (blue line) bias.
Dashed lines are the power law fits assuming $T_{\rm bath}\ll T_C$.
(d) Differential conductances of the suspended SGS junction at negative bias
measured at $T_{\rm bath}=10$mK (superconducting leads, black line) and at $T_{\rm bath}=1$K (normal leads, blue line). 
Red dashed line shows the fit with Eq. (\ref{asym}) and the arrow points to the MAR peak corresponding to the voltage $\Delta/e$. }
\label{plot1}
\end{figure*}

The superconducting-to-normal transition due to Joule heating in a suspended bilayer SGS sample is depicted in Fig. \ref{heating}a for a temperature range of $10\dots 800$\,mK ($T_C \simeq 770$\,mK). Each resistance peak is clearly split into two parts, presumably because of junction asymmetry.  There is also a small asymmetry between positive and negative bias voltages, giving rise to a total of four peaks. 

Heat flow in mesoscopic devices can typically be characterized by a power law of the form  
\begin{equation}
P=\Sigma (T_{L,R}^\gamma - T_{\rm bath}^\gamma)
\label{PS1}
\end{equation}
where $\Sigma$ is a  constant and $\gamma$ is a characteristic exponent which depends on the dissipation mechanism. The transition to the normal state occurs at the critical temperature $T_{L,R} = T_C$ and the corresponding critical powers can be expressed as
\begin{equation}
P_C = \frac{P_{C,0}}{T_C^\gamma} (T_C^\gamma - T_{\rm bath}^\gamma),
\label{PC}
\end{equation}
where $P_{C,0}$ is the critical power at $T_{\rm bath} = 0$. 
The observed peak positions from Fig. \ref{heating}(a) are plotted in Fig. \ref{heating}(b), along with the fits using Eq. (\ref{PC}) 
and the exponent $\gamma$ as a fit parameter. We obtain $\gamma\approx 3.1$ for all four peaks. 
The four resistance traces in  Fig. \ref{heating}(b) converge to two diﬀerent critical temperatures, $\sim 760$\,mK and $\sim 780$\,mK.
This supports the idea that the two leads have slightly different critical temperatures.

Next, we analyze the dependence of the differential conductance of the suspended SGS junction on bias voltage.
First, we find bias dependent excess current by subtracting 
the IV-curves of the junction measured at the base temperature of $T = 10$ mK, when the leads are superconducting, and at 1K, when the leads become normal,
i.e. we define $I_{\rm exc}(V) = I_{\rm 10mK}(V)-I_{\rm 1K}(V)$.
This current is plotted in Fig. \ref{plot1}(a).
Two red dashes lines show the maximum and the minimum value of $I_{\rm exc}(V)$, they are equal to $\pm 26.5$ nA
and should correspond to the non-suppressed theoretical excess current $\pm I_{\rm exc}$ defined in Eq. (\ref{I(V)}).
Analyzing multiple Andreev reflection pattern in $dI/dV$ for this sample (see Fig. \ref{plot1}(d)),
we estimate the superconducting gaps $\Delta_L\approx \Delta_R=90$ $\mu$eV and find the junction resistance  $R=2195$ $\Omega$. 
With these values, we obtain the excess current suppression parameter from Eq. (\ref{I(V)}) as $\alpha=0.646$. 
From Eq. (\ref{diff}) we estimate the ratio $L_{\rm jct}/\xi \approx 3.8$ and $\xi\approx 130$ nm.
The latter value is similar to the one obtained for the non-suspended junction, which supports the validity
of the diffusive model for graphene.

In Fig. \ref{plot1}(b) we plot the effective temperature of the leads $T_0(V)$, which is obtained from the bias dependent excess current
plotted in Fig. \ref{plot1}(a)
by numerically solving the equation $|I_{\rm exc}(V)|/I_{\rm exc}^{\max} = \Delta(T_0)/\Delta_0$ and
assuming BCS gap-temperature dependence, $\Delta(T_0)\approx\Delta_0\tanh(1.737\sqrt{T_C/T_0-1})$.
We used $T_C=770$ mK, which is the average value between the critical temperatures of the two leads given above.
The accuracy of this procedure is limited by the noise in $I_{\rm exc}(V)$, which stems from the
numerical subtraction of the two  experimental I-V curves. 
For this reason, the splitting between the two critical voltages
$V_{CL}$ and $V_{CR}$, which is clear in Fig. \ref{heating}(a), is not visible in Fig. \ref{plot1}(a). 
Therefore, here we ignore relatively weak asymetry between the leads and put $T_L=T_R=T_0$.  
Furthermore, this type of thermometry only makes sense in the limited range of bias voltages where
the excess current decreases with the bias, but the critical temperature of the leads is not yet reached. 
Therefore, only these two voltage intervals at positive and at negative bias are shown in Fig. \ref{plot1}(b).
The dependence $T_0(V)$, obtained in this way, resembles the model prediction
shown in Fig. \ref{plots}(a).

In Fig. \ref{plot1}(c) we plot the Joule heating power $P=IV$, 
which is equal to the cooling power of the junction, versus the temperature of the leads $T_0(V)$ 
presented in Fig. \ref{plot1}(b).
The red line corresponds to the positive bias and the blue one -- to the negative bias. 
The dashed lines present the predictions of Eq. (\ref{PS1}) at very low bath temperature, $T_{\rm bath}\to 0$, 
and with the same values of the exponent $\gamma=3.1$ and $\Sigma=P_{C,0}/T_C^\gamma$ as in the theory curves
shown in Fig. \ref{heating}(b). Namely, we have used $P_{C,0}=4.7$ nW for positive and $P_{C,0}=4.4$ nW for negative bias,
and these values match those extracted from the positions of the peaks in $dV/dI$ with larger power, see Fig. 5(b).
The agreement between the dashed lines and the experimental curves is good at relatively high temperatures close to $T_C$.
Thus, we have confirmed the validity of Eq. (\ref{PS1}) in a broader range of temperatures by varying the temperature of
the leads $T_0(V)$ with the bias voltage and keeping the bath temperature temperature low, $T_{\rm bath}\sim 10$ mK.
At  temperatures $T_0(V)\lesssim 0.6$ K the power law curves deviate from the experimetal data. However, there
the accuracy of the numerical conversion of the excess current to temperature becomes very poor and we cannot draw any
conclusion from this discrepancy.

The differential conductances of the suspended SGS junction in the superconducting ($T_{\rm bath}=10$ mK, black line) and 
in the normal ($T_{\rm bath}=1$ K, blue line) states are presented in Fig. \ref{plot1}(d). 
We show only negative bias voltages, where the dips
associated with the transition of the leads to the normal state are more prononced.
In the same figure we also show the fit based on Eq. (\ref{asym}) with the red dashed line. 
The theoretical curve has been plotted with the paramters earlier extracted from the fits shown in Figs. \ref{heating}(b) and \ref{plot1}(a),
namely, we have used $\alpha=0.646$, $R=2195$ $\Omega$ and $P_{C,0}=4.4$ nW. With these parameters,
we obtain the thermal conductance at critical temperature $G_C^{\rm th}=\gamma P_{C,0}/T_C=17.7$ nW/K. To obtain this value, 
we have taken the derivative of the cooling power (\ref{PS1})
at $T_{\rm bath}=T_C$ and, as before, have set $\gamma=3.1$ and $T_C=0.77$ K.
The parameter $b$, defined in Eq. (\ref{alpha}), takes the value $b=2.776\times 10^4$.
The two critical voltages were adjustable parameters, they were chosen as  $V_{CL}=-2.82$ mV and $V_{CR}=-3.2$ mV.
In addition, we also introduced the non-linear correction to the differential conductance of the junction in the normal state,
which originated from the energy dependent density of states in graphene and
was needed to better reproduce the voltage dependence of the blue line in Fig. \ref{plot1}(d). Accordingly, in Eq.  
(\ref{asym}) we have replaced $1/R \to (1+\beta V)/R$ with  $\beta=0.015$ mV$^{-1}$.
We find that Eq. (\ref{asym}), modified in this way, rather well reproduces the shape of the experimental $dI/dV$-curve 
everywhere except at low bias voltages, where the MAR peaks become visible.
As expected, the dips caused by transistion of the leads to the normal state 
are smeared in the experimental curve. 
We also note that Fig. \ref{plot1}(d) is a different representation of the inset in Fig. \ref{heating}(b),
and that it resembles the theoretical model plot of Fig. \ref{plots}(c).

We conclude this section by noticing that the model of Sec. \ref{theory}, 
which relates the dips in the differential conductance of a highly transparent Josephson junction to the suppression
of the excess current by Joule heating, well describes our experimental data.

\section{Discussion and conclusions}
\label{conclusion}

In the previous section we have demonstrated how one can obtain the dependence of the cooling power 
of the Josephson junction on the bath temperature and on the junction temperature by fitting
its differential conductance at high bias voltages above $2\Delta/e$. 
This simple technique is applicable to any highly transparent Josephson junction and it
might be useful for high-flux calibration of bolometers containing such junctions.   

Our experimental results on graphene deserve further specific discussion.
We find that the critical Joule heating $P_C$  becomes independent of the bath temperature at $T_{\rm bath} \ll T_C$,
see Fig. \ref{heating}(b). 
Such saturation of the heat flow at low temperature 
is rather common in systems with weak coupling between electrons and lattice phonons. 
We find similar behavior in the theoretical electronic heat transport model of Sec. \ref{suspended}. 
Indeed, for $T_0=T_C$ and at $T_{\rm bath} \ll T_C$, where the energy gap of the superconductor does not depend on temperature, the 
heat power (\ref{PSfull}) saturates at the limiting value (\ref{Pc}).

According to Figs. \ref{NS_IV}(a) and \ref{heating}(b), the saturated low-temperature critical power amounts to $P_C^{\mathrm{NS}}=1$\,nW and $P_C^{\mathrm{S}}=4$\,nW in our non-suspended and suspended sample, respectively. If we scale these values by the number of leads (see Table I) and the cross sectional area of the sandwich conductors, $d \times W_{\mathrm{leads}}$, 
we obtain the values  $P_C^{\mathrm{NS}}/(d \times W_{\mathrm{leads}}) = 8.67$\,kW/m$^2$ 
and $P_C^{\mathrm{S}}/(d \times W_{\mathrm{leads}}) =13.7$\,kW/m$^2$. 
The similarity of these two critical power densities supports 
the conclusion that thermal transport along the leads is the main mechanism of heat relaxation in our devices.

The critical voltages amount to $V_C^{\mathrm{NS}} = 0.34$\,mV and $V_C^{\mathrm{S}} = 3.0$\,mV for samples NS and S, respectively. 
We have applied Eq. (\ref{Vc}) at low $T$ to investigate whether these values scale according to the  model. 
Since we do not know exactly the lead resistance, we compare the measured ratio of critical voltages to the theory by 
assuming that the lead resistance scales inversely proportional to the cross sectional area of the leads \cite{footnote}. 
With the parameters from Table I, Eq. (\ref{Vc}) yields for the critical voltage ratio $V_C^{\mathrm{S}}\large{/}V_C^{\mathrm{NS}}=8.5$, 
while from the experimental data in Figs. \ref{NS_IV} and \ref{heating} we obtain $V_C^{\mathrm{S}}\large{/}V_C^{\mathrm{NS}}=8.8$. 
Thus, our theoretical model accounts for the variation of the critical voltage within a range covering one order of magnitude.

The observed exponent in the temperature dependence of the power for the suspended junction, 
$\gamma = 3.1$, is close to the one found by Borzenets {\it et al} for SGS junctions 
with Pb leads and graphene deposited on Si/SiO$_2$ substrate \cite{Borzenetz2013}.
Lead has a large superconducting gap, which should effectively cut off electronic thermal conductance below 1 K and 
leave only electron-phonon coupling for energy relaxation \cite{Voutilainen2011}.
For SGS junctions with Al/Ti leads on the Si/SiO$_2$ substrate the exponents $\gamma\approx 4$ \cite{Choi}
and $\gamma=3.1$ \cite{Tsumura}, which agrees with our findings, have been previously reported. 
Power law thermal transport with the exponent $\gamma=4.5$   has also been observed 
in superconducting Nb at low temperatures $k_BT\ll\Delta$ \cite{Feshchenko2017}.

The origin of the exponent $\gamma\approx 3$ observed in our suspended device is not fully clear.
Let us briefly discuss several possible explanations for this observation.
In our experiment the ratio $k_B T/\Delta$ extends to large values, 
and, therefore, energy relaxation should be dominated by quasiparticles carrying heat through the leads. 
For temperatures close to $T_C$ the quasiparticles behave almost as
normal state electrons obeying Wiedemann-Franz law, which leads to $\gamma=2$, see Eq. (\ref{PS_high}).
The increase of the exponent $\gamma$ from 2 to 3
may be qualitatively explained by several different effects or by their combination. 
First, the quasiparticle heat current  (\ref{PSfull})
is lower and more sensitive to temperature than the normal state one in the whole interval $T<T_C$, including the temperatures close to $T_C$
as Eq. (\ref{PS_high}) shows. 
Second, the parameter $R_{\rm leads}$, appearing in the cooling power (\ref{PSfull}),
is the resistance of the part of an Al lead in which thermalization between the quasiparticles and the substrate phonons occurs.
Since the corresponding thermalization length $L_T$ is reduced with increasing temperature, 
the resistance $R_{\rm leads}\propto L_T$ is also reduced. Hence,
the dependence of the power (\ref{PSfull}) on temperature becomes stronger than it were in the model with fixed $R_{\rm leads}$.
Yet another possibility for the increase of the exponent
is the injection of strongly non-equilibrium quasiparticles with non-thermal distribution
in the Al leads at high bias voltages. The propagation of such quasiparticles should be
described by the full set of kinetic equations with the inclusion of charge imbalance instead of
just one Eq. (\ref{balance_equation}).  
Quantitative analysis of the effects mentioned above goes beyond the scope of this paper.
At the moment, the similarity of the exponents observed in our experiment and in Refs. \cite{Borzenetz2013,Tsumura} remains puzzling.

In conclusion, we have analyzed, both theoretically and experimentally, the effect of Joule heating on the IV characteristics 
of highly transparent Josephson junctions. 
We have shown that the Joule heating supresses the excess current
and induces dips in the differential conductance at bias voltages where
the two leads switch to the normal state. We have derived simple analytical expressions for
the shape of these dips and for their positions in a specific case of a suspended junction, which
is cooled only through the leads.
We have experimentally tested preditions of the theory by studing SGS junctions
in both suspended and non-suspended configurations. 
Good agreement between the experiment and the theory is found. 
In the suspended junction we observe power law temperature dependence of the heat current with the exponent
$\gamma=3.1$ instead of theoretically expected exponential dependence (\ref{PS_low}) at low temperatures and
power law scaling with $\gamma=2$ at higher temperatures (\ref{PS_high}). 
Our work also indicates that excess current 
can be employed for thermometry in special cases.

\section*{Acknowledgements}
 Discussions with Manohar Kumar and Alexander Savin are gratefully acknowledged. This work was supported by the Academy of Finland projects 314448 (BOLOSE), 310086 (LTnoise) and 312295 (CoE, Quantum Technology Finland) as well as by ERC (grant no. 670743). This research project utilized the Aalto University OtaNano/LTL infrastructure which is part of European Microkelvin Platform (funded by European Union’s Horizon 2020 Research and Innovation Programme Grant No. 824109). A.L.\ is grateful to V{\"a}is{\"a}l{\"a} foundation of the Finnish Academy of Science and Letters for scholarship.
 The work at HSE University (M.R.S. and A.S.V.) was supported by the framework of the Academic Fund Program at HSE University in 2021 (grant No 21-04-041).



\begin{thebibliography}{99}
\bibitem{footnote}
The normal state resistance of the Al layer is much smaller than that of the Ti layer. Hence, we also checked the scaling using the Al layer thickness, but the difference is small, and the change would not modify our conclusions. We use the total thickness because the mechanism of the electronic heat transport in the superconducting state is not fully clear.
\end{thebibliography}
\end{document}